\documentclass[12pt]{article} 
\title{Poincar\'{e} Connections in Flat Spacetime}  
\author{{\it Richard Shurtleff~}\thanks{affiliation and mailing 
address: Department of Applied Mathematics and Sciences, 
Wentworth Institute of Technology, 550 Huntington Avenue, 
Boston, MA, USA, ZIP 02115, telephone number: (617) 989-4338, fax 
number: (617) 989-4591 , web: http://ox.wit.edu//~shurtleffr/ , e-mail address: shurtleffr@wit.edu}} 
\begin{document} 
          
\maketitle 

\begin{abstract} In flat spacetime, as a simple 4-vector, a particle's 4-velocity cannot be changed by translation. Parallel translation then produces constant velocity, motion without force. Here we consider a richer, but less well-known, representation of the Poincar\'{e} group of symmetries of flat spacetime in which translation adds a non-linear connection term to the 4-vector. Then 4-vectors can be meaningfully parallel translated, curved geodesics can be developed, and the curvature for the connection can be found. By adding an assumption, that the arc length is calculated with a position dependent metric, it is shown that a geodesic has a 4-acceleration which is the sum of a Christoffel connection term and a term that is the scalar product of 4-velocity and an antisymmetric tensor. This is just like the force law obeyed by a charged, massive particle in general relativity. Thus the dynamical laws of electrodynamics and general relativity can be deduced as geodesic equations from the way particle proper time depends on position and the way 4-vectors can change upon translation in flat spacetime.

\end{abstract}

\pagebreak
\section{Introduction}

Consider a 4-vector such as a particle's velocity and a 2nd rank tensor such as the electromagnetic field tensor. Each determines a representation of the homogeneous Lorentz group of rotations and boosts in flat spacetime. But neither vector nor tensor changes upon translation, which means that all translations are represented by the identity transformation. Thus, the Poincar\'{e} group, which has all the transformations of the Lorentz group as well as translations, has representations determined by vectors and tensors, but these representations represent translations unfaithfully.

Now suppose a field $\psi$ is created as the direct sum of a velocity field and a tensor field, 
$$ \psi = \pmatrix{v^{\mu} \cr T^{\alpha \beta}_{0}} \quad ,
$$
where the greek indices range from 1 to 4, with 4 being the time index. The field $\psi$ determines representations of the Lorentz group that, not surprisingly, are simply the application of the vector transformations to the vector part and the tensor transformations applied to the tensor part. Perhaps unexpectedly, however, translations do change $\psi.$ Of course translations need not change $\psi,$ but that is only a special case. In general translations are represented faithfully.  

An example of such a combination is given by Minkowski coordinates $x^{\mu}.$  Under rotations about the origin and boosts, the transformations of $x^{\mu}$ are identical to the transformations of 4-vectors. For a Lorentz transformation $\Lambda,$ coordinates change to ${x^{\mu}}^{\prime}$ = $\Lambda^{\mu}_{\nu} x^{\nu},$ with summation over repeated indices. But coordinates also change upon translation, ${x^{\mu}}^{\prime}$ = $\Lambda^{\mu}_{\nu} x^{\nu} + \delta x^{\mu}.$ The key to this is the fact that coordinates transform like the direct sum of a vector field with a trivial scalar field.\cite{Hamermesh}
$$ \pmatrix{x^{\mu} \cr 1}^{\prime} = \pmatrix{\Lambda^{\mu}_{\nu} && \delta x^{\mu} \cr 0 && 1}\pmatrix{x^{\nu} \cr 1} = \pmatrix{\Lambda^{\mu}_{\nu} x^{\nu} + \delta x^{\mu} \cr 1} \quad .
$$
The direct sum has translation dependence whereas neither the vector part by itself nor the scalar part by itself has any translation dependence.

Technically, the general rule for non-trivial translation matrices for spin $(A,B)\oplus(C,D)$ is that $\mid A-C \mid$ = $\mid B-D \mid$ = 1/2.\cite{Shurtleff} Since 4-vectors transform as spin $(1/2,1/2),$ there are just four options: $(1/2,1/2)\oplus(0,0),$ $(1/2,1/2)\oplus(0,1),$ $(1/2,1/2)\oplus(1,0),$ and $(1/2,1/2)\oplus(1,1).$ And spins $(0,0),$ $(0,1),$ $(1,0),$ and $(1,1)$ combine to make the transformations of a 2nd rank tensor. Since all four spin options are included with the transformations of a 2nd rank tensor, the most general translation matrices for transforming 4-vectors are produced by the direct sum of a 4-vector field and a 2nd rank tensor field.

Thus, upon translation, the change in the 4-vector field in $\psi$ acquires a {\it{Poincar\'{e} connection}} term. The term is not necessarily proportional to the vector field, so the connection is in general a `non-linear'  connection. The usual ideas can be applied with this connection and are applied here in Section \ref{Parallel}. It is possible for a vector field to change along a curve in just the right way necessary for translations along the curve to leave the vector field invariant. Such a vector field is said to be {\it{parallel translated}} along the curve. And a {\it{geodesic curve}} has a tangent vector that is parallel translated along the curve. Also, round trip parallel transport tests for {\it{curvature}}. (The term `geodesic' curve is used here, as it often is elsewhere, to mean `parallel translated tangent' curve and not to indicate extreme arc length.) 

The tangent vector to a curve depends not only on what events make up the curve, but also on how distance along the curve is calculated. In symbols, $dX^{\mu}/d\tau$ depends on $d\tau$. When the arc length $\tau$ is calculated with a metric $g_{\alpha \beta},$ the tangent vector has a constant magnitude. This follows  because dividing the equation $(d\tau)^2 = -g_{\alpha \beta} dX^{\alpha} dX^{\beta}$ by $(d\tau)^2$ implies the magnitude squared, $g_{\alpha \beta} (dX^{\alpha}/d\tau) (dX^{\beta}/d\tau),$ is a constant. 

The tangent vector of a geodesic is already constrained by the geodesic equation. To further constrain the tangent vector by requiring it to have constant magnitude puts constraints on the connection. Since the connection is directly related to the tensor part of $\psi,$ only special tensor fields produce geodesics that are parametized by an arc length calculated with a metric. 

Suppose the metric is the constant metric of flat spacetime. Then, as shown in Section~\ref{LorentzForce}, the geodesic equation together with constant tangent vector magnitude requires an antisymmetric tensor, $T^{\beta \alpha}_{0}$ = $-T^{\alpha \beta}_{0}.$ And the geodesic equation for an antisymmetric tensor is the same as the equation of motion for a charged, massive particle in an electromagnetic field when the tensor field is proportional to the electromagnetic field, $k T^{\alpha \beta}_{0}$ = $-(e/m)F^{\alpha \beta},$ where $e$ and $m$ are the charge and mass of the particle and $k$ is a constant.  (For selected values of the arbitrary constants that appear in the translation matrices, the connection is proportional to the 2nd rank tensor field accompanying the 4-vector field in $\psi.$) Thus, the field $\psi$ with particle velocity as the vector part and the charge to mass ratio times the electromagnetic field as the tensor part produces geodesics that are the paths of charged particles. 

However one can deduce from observations that particles behave as if their proper time depends on distance from sources of gravitation.\cite{Will} Clocks at a higher altitude run faster than clocks at a lower altitude, an effect that also depends on speed. Hence a more appropriate arc length for particle motion should depend on a metric $g_{\alpha \beta}$ that is a function of location. It is shown in Section \ref{Gravity} that the dual requirements of the geodesic equation and the constant magnitude associated with a position dependent metric now requires the tensor field to include an affine connection term, $k T^{\sigma}_{(0) \alpha}$ = $-(e/m)F^{\sigma}_{\alpha} +$ $ \Gamma^{\sigma}_{\alpha \gamma} (dX^{\gamma}/d\tau).$ The geodesic equation is then the equation of motion of a charged, massive particle in an electromagnetic and gravitational field when the metric $g_{\alpha \beta}$ is the appropriate gravitational metric.     

The geodesic equations obtained in this article are appropriate for particle dynamics, but it should be emphasized that this article does not include any attempt to determine the field equations obeyed by the electromagnetic field $F^{\mu \nu}$ and the gravitational metric $g_{\alpha \beta}.$ Having both the symmetries of the flat spacetime metric and the position dependent gravitational metric for calculating particle proper times makes this a {\it{bimetric}} formalism, by definition and there are well known bimetric theories \cite{bimetric1,bimetric2,bimetric3} that propose alternatives to general relativity. None of these alternatives are considered here.

In this article, it is assumed for definiteness and by the success of its experimental predictions that the field equations of general relativity determines the position dependent metric. In a given reference frame of the flat spacetime, each event has unique coordinates, the gravitational metric has some value at each event, geodesics can be calculated as functions of the flat spacetime coordinates, and the motion of falling particles may be determined. 

By parallel translation of a 4-vector around a closed path, one finds an expression for the round trip deviation that depends on the Riemann-Christoffel curvature tensor of the gravitational metric and the covariant derivatives of the electromagnetic field. Since the spacetime upon which these quantities is constructed is flat the curvature is not the curvature of that flat spacetime. Rather the curvature is the curvature of the  Poincar\'{e} connection  $T^{\mu}_{\nu}.$ It is not unusual in mathematical treatments of differential geometry to define the curvature of a connection. So the curvature found in Section \ref{Gravity} is not that of flat spacetime, but of a connection developed in flat spacetime that is associated with the position dependent metric.

If the proper time of all matter is calculated with the position dependent gravitational metric $g_{\alpha \beta},$ then surveying flat spacetime might not be possible. The sequence of measurements to determine the flat spacetime coordinates $x^{\mu}$ of an event, i.e. an inertial frame, assumes that rulers and clocks move freely. If the equivalence principle is assumed, then gravitation is a universal force and there are no freely moving rulers and clocks, even in principle. If there are no clocks that measure the time of flat spacetime, is the term `flat spacetime' meaningful? This quandary is discussed in Section \ref{Survey}. 

Perhaps the realization that a flat spacetime exists whose symmetries can be the basis for motion in gravitational and electromagnetic fields is the one benefit of this analysis. As indirect evidence, one can point to the observed paths followed by matter in electromagnetic and gravitational fields. After all, the observed paths are much like the geodesics that are determined by ({\it{i}}) the symmetries of that flat spacetime and ({\it{ii}}) the assumption of a metric based arc length.

 \pagebreak
\section{Transformations of the Direct Sum Field $\psi$ } \label{psi}

By repeatedly observing the motion of a particle in a given force field, the possible paths in the force field can be determined. Thus the 4-velocities along the various paths form 4-vector fields. If the fields are to be single-valued, then it is necessary to select non intersecting particle paths for any one field. To avoid intersecting paths and keep the field single-valued, it may also be necessary to restrict the domain of any one field to a 4-dimensional patch of events in spacetime. 

The transformations of a field of 4-velocities follows the rules for transforming fields of 4-vectors. In this section, the group of transformations considered is made up of successive infinitesimal rotations about the origin, boosts and translations, i.e. the symmetry transformations that preserve the flat spacetime metric and are connected to the identity transformation. 

The most general such transformations of a 4-vector field occur when the 4-vector field is accompanied by a tensor field. The direct sum of a 4-vector field $v^{\mu}(x)$ and a 2nd rank tensor field $T^{\alpha \beta}_{0}(x)$ can be written as a column vector,
\begin{equation} \label{psi1}
 \psi_{l}(x)  = \pmatrix{v^{\mu}(x) \cr T^{\alpha \beta}_{0}(x)} \quad ,
\end{equation}
where, for convenience, the vector and tensor are assumed to be dimensionless and the index $l$ takes values in the set $\{1, \ldots , 20 \}$ since the vector has 4 components and the tensor has 16 components. Transcribing the 16 double index values $\alpha \beta$ = $\{11,12,\ldots, 44\}$ to the 16 single index values $l$ = $\{5,6,\ldots,20\}$ is treated in the Problem set, Appendix \ref{A}.

Any Poincar\'{e} transformation $\{\Lambda,\delta x\}$ can be written as a homogeneous Lorentz transformation $\Lambda$ followed by a translation through a displacement $\delta x.$ The field $\psi$ changes as a differential representation that changes functions of $x$ to functions of $\Lambda x + \delta x$ and also as a square matrix representation that acts on the 20 components of $\psi_{l},$ \cite{Tung1, WeinbergFields}
\begin{equation} \label{transf2}
U(\Lambda,\delta x) \psi_{l}(x) {U}^{-1}(\Lambda,\delta x) = \sum_{\bar{l}} D^{-1}_{l \bar{l}}(\Lambda,\delta x)  \psi_{\bar{l}}(\Lambda x + \delta x) \quad ,
\end{equation}
where $D(\Lambda,\delta x)$ is the covariant nonunitary matrix representing the spacetime transformation $\{\Lambda,\delta x\}.$ 

Let $J^{\mu \nu}$ be the angular momentum and boost matrix generators of rotations and boosts and let $P^{\mu}$ be the four momentum matrices. Then the transformation matrices are determined by the displacement $\delta x_{\mu}$ and the antisymmetric parameters $ \omega_{\mu \nu}$ of the Lorentz transformation $\Lambda,$
\begin{equation} \label{transf1}
 D(\Lambda,\delta x_{\mu})  = \exp{(- i \delta x_{\mu} P^{\mu})}\exp{( i \omega_{\mu \nu} J^{\mu \nu}/2)}  \quad .
\end{equation}

The generators can be put in block matrix form with angular momentum and boost generators in the form
\begin{equation} \label{gen1}
 J^{\rho \sigma}  = \pmatrix{(J^{\rho \sigma}_{11})^{\mu}_{\nu} && 0 \cr 0 && (J^{\rho \sigma}_{22})^{\alpha \beta}_{\gamma \delta}} \quad ,
\end{equation}
where the 11-block generators are herein taken to be 
$$ 
 (J^{\rho \sigma}_{11})^{\mu}_{\nu}  = i \left(\eta^{\sigma \mu} \delta^{\rho}_{\nu} - \eta^{\rho \mu} \delta^{\sigma}_{\nu} \right) \quad ,
$$ 
where $\eta^{\alpha \beta}$ is the flat spacetime metric, the diagonal matrix $\eta$ = diag$\{+1,+1,+1,-1\}$ and $\delta^{\rho}_{\nu}$ is the Kronecker delta function, one for $\rho$ = $\nu$ and zero otherwise. The 22-block generators are
$$ 
 (J^{\rho \sigma}_{22})^{\gamma \delta}_{\epsilon \xi}  = -i \left(\eta^{\rho \gamma} \delta^{\sigma}_{\epsilon} \delta^{\delta}_{\xi}- \eta^{\sigma \gamma} \delta^{\rho}_{\epsilon} \delta^{\delta}_{\xi} + \eta^{\rho \delta} \delta^{\sigma}_{\xi} \delta^{\gamma}_{\epsilon}- \eta^{\sigma \delta} \delta^{\rho}_{\xi} \delta^{\gamma}_{\epsilon} \right) \quad .
$$ 
The formulas for $J^{\mu \nu}_{11}$ and $J^{\mu \nu}_{22}$ are standard formulas, taken from the literature.\cite{WeinbergJ}

There are two sets of momentum matrices, those with the 12-block nonzero and those with the 21-block nonzero. The 21-block momentum matrices are not displayed here because the 12-block matrices are needed. One finds that
\begin{equation} \label{gen4}
 P^{\mu}  = \pmatrix{0 && (P^{\mu}_{12})^{\nu}_{\alpha \beta} \cr 0 && 0} \quad .
\end{equation}
Obviously, from their block matrix form, the matrix product of two momentum matrices vanishes, $P^{\mu} P^{\nu}$ = 0. Expressions for the components of momentum matrices needed here are not readily available from the literature. They can be obtained by solving the Poincar\'{e} algebra given the expressions above for $J^{\mu \nu}$ or from the formulas in reference \cite{Shurtleff}. A convenient formula for the 12-block components is given by 
$$ 
 (P^{\mu}_{12})^{\nu}_{\alpha \beta}  = i [ \left( C_1 + C_2 - C_3 \right)\delta^{\mu}_{\alpha} \delta^{\nu}_{\beta}  + 
 \left( C_1 + C_2 + C_3 \right)\delta^{\mu}_{\beta} \delta^{\nu}_{\alpha}  -  2 C_1 \eta^{\mu \nu} \eta_{\alpha \beta}
  + C_4  \eta^{\mu \sigma} \eta^{\nu \kappa} \epsilon_{\sigma \kappa \alpha \beta}]  \quad ,
$$ 
where the constants $C_{i}$ have dimensions of an inverse distance and can be chosen arbitrarily. These constants appear because the Poincar\'{e} algebra is homogeneous in $P^{\mu},$ so the momentum matrices are determined within a scale factor. There are four such constants because the transformations of a 2nd rank tensor combine four spins. 

The generators $J^{\mu \nu}$ and $P^{\mu}$ satisfy the Poincar\'{e} algebra. The proof is relegated to the problem set, see Appendix \ref{A}. It can also be shown that the matrices $D(\Lambda,\delta x)$ in (\ref{transf1}) with these generators form a representation of the Poincar\'{e} group.

\pagebreak
\section{Translation, Parallel Translation, and Geodesics } \label{Parallel}

While the formalism described in Section \ref{psi} includes boosts and rotations, this article is focused on translations, since it is the translations that lead to connection terms for the 4-vector part of $\psi.$

The general expressions in Section \ref{psi} are greatly reduced with simple translations. The homogeneous Lorentz transformation is just the identity transformation, $ \Lambda$ = 1 and $ \omega_{\mu \nu}$ = 0. Also, the translation matrix $D(1,\delta x_{\mu})$ simplifies because the matrix product of two momentum matrices vanishes, eliminating terms higher than first order, $D(1,\delta x_{\mu})$ = $\exp{(- i \delta x_{\mu} P^{\mu})}$ = $(1 + i \delta x_{\mu} P^{\mu}).$ And one can assume that $\delta x_{\mu}$ is small. 

Applying these simplifications leads to the following sequence of expressions
$$ 
U(1,\delta x) \psi_{l}(x) {U}^{-1}(1,\delta x) = D^{-1}_{l \bar{l}}(1,\delta x)  \psi_{\bar{l}}( x + \delta x) 
$$ $$
 =  (1 + i \delta x_{\mu} P^{\mu})_{l \bar{l}} \enspace \psi_{\bar{l}}( x + \delta x) 
$$ $$
 = \psi_{l}( x + \delta x) +   i \eta_{\mu \sigma} \delta x^{\sigma} P^{\mu}_{l \bar{l}} \psi_{\bar{l}}( x + \delta x)  
$$ $$
 = \psi_{l}(x) +  \left[ \frac{\partial{ \psi_{l}(x)}}{\partial{x^{\sigma}}} + i \eta_{\mu \sigma} P^{\mu}_{l \bar{l}} \psi_{\bar{l}}(x)\right] \delta x^{\sigma}+  {\mathrm{O}}[(\delta x)^2]\quad .
$$
Thus the field $\psi$ changes by $\delta \psi_{l},$
$$
\delta \psi_{l} = \left[ \frac{\partial{ \psi_{l}(x)}}{\partial{x^{\sigma}}} + i \eta_{\mu \sigma} P^{\mu}_{l \bar{l}} \psi_{\bar{l}}(x)\right] \delta x^{\sigma} +  {\mathrm{O}}[(\delta x)^2] 
$$ $$
= \pmatrix{\left[ \frac{\partial{ v^{\nu}(x)}}{\partial{x^{\sigma}}} + i \eta_{\mu \sigma} (P^{\mu}_{12})^{\nu}_{\alpha \beta} T^{\alpha \beta}_{0}(x) \right] \delta x^{\sigma} +  {\mathrm{O}}[(\delta x)^2] \cr T^{\alpha \beta}_{0}(x + \delta x)}   \quad .
$$
By isolating the 4-vector part of the change in $\psi,$ one finds that the translation has changed the 4-vector part of $\psi$ by an amount $\delta v^{\nu}(x)$ given to first order by 
\begin{equation} \label{Dpsi3}
\delta v^{\nu} =  \left( \frac{\partial{ v^{\nu}}}{\partial{x^{\sigma}}} - T^{\nu}_{\sigma} \right) \delta x^{\sigma} \quad ,
\end{equation}
where 
\begin{equation} \label{Dpsi4}
T^{\nu}_{\sigma} = \eta_{\mu \sigma} T^{\mu \nu} =  -i \eta_{\mu \sigma} (P^{\mu}_{12})^{\nu}_{\alpha \beta} T^{\alpha \beta}_{0}  \quad ,
\end{equation}
and $\delta v^{\nu},$ $ v^{\nu},$ and $T^{\nu}_{\sigma}$ are understood to be functions of event coordinates $x^{\mu}.$ 

The quantities $T^{\nu}_{\sigma}$ and $T^{\mu \nu}$ are herein called {\it{Poincar\'{e} connection tensors}} because they appear as an additive term to a partial derivative in (\ref{Dpsi3}) and because they are 2nd rank tensors for the homogeneous Lorentz transformations generated by the $J^{\mu \nu}_{22}$ matrices in (\ref{gen1}).

The two terms in parentheses in (\ref{Dpsi3}) represent two different Poincar\'{e} representations. The partial derivative appears because of the differential representation of the Poincar\'{e} group and the connection term arises from the 20-dimensional nonunitary representation of the Poincar\'{e} group in (\ref{gen1}) and (\ref{gen4}). Note that for certain values of the constants in (\ref{gen4}), the connection tensor is proportional to the tensor part of $\psi,$ i.e. $T^{\alpha \beta}$ = $ k T^{\alpha \beta}_{0}$ for $C_{2}$ = $-C_{3}$ = $k/2$ and $C_{1}$ = $C_{4}$ = 0, where $k$ is a constant with the dimensions of an inverse length.

It is clear from (\ref{Dpsi3}) that if the field is invariant to translation in any direction, the change in the 4-vector field vanishes for arbitrary small displacements. When  $\delta v^{\nu}(x) $ = 0 with $\delta x^{\sigma}$ arbitrary one has
$$
 \frac{\partial{ v^{\nu}}}{\partial{x^{\sigma}}} = T^{\nu}_{\sigma}   \quad .
$$
This constrains $T$ because the second partials of $v^{\nu}(x)$ commute,
$$
 \frac{\partial^2{ v^{\nu}}}{\partial{x^{\rho}}{\partial{x^{\sigma}}}} = \frac{\partial^2{ v^{\nu}}}{\partial{x^{\sigma}}{\partial{x^{\rho}}}}   
$$
and it follows that
$$
\frac{\partial{ T^{\nu}_{\sigma}}}{\partial{x^{\rho}}} - \frac{\partial{ T^{\nu}_{\rho}}}{\partial{x^{\sigma}}} = 0  \quad .
$$

Theorem: If a vector field is invariant to translations in any direction, then the `curl' of the connection tensor must vanish. 

Next, consider translation along a suitably differentiable curve $X(\tau).$ Let $\tau$ for now be any suitably differentiable arc length along the curve. The displacement $\delta x^{\sigma}$ is restricted to being tangent to the curve; one has
$$
 \delta x^{\sigma} = \frac{dX^{\sigma}}{d\tau} \delta \tau   \quad .
$$
Also, define the vector field restricted to the curve to be
$$
 V^{\mu}(\tau) \equiv v^{\mu}(X(\tau))     \quad ,
$$
which implies
$$
  \frac{dV^{\mu}}{d\tau} = \frac{\partial{ v^{\mu}}}{\partial{x^{\sigma}}}\frac{dX^{\sigma}}{d\tau}   \quad .
$$
By (\ref{Dpsi3}), translation along a short segment of the curve changes the vector field by $\delta V^{\nu}(\tau)$, where 
$$
 \delta V^{\mu} =   \left[ \left(\frac{\partial{ v^{\mu}}}{\partial{x^{\sigma}}}\right) -  T^{\mu}_{\sigma}\right] \frac{dX^{\sigma}}{d\tau}  \delta \tau \quad ,
$$
which implies that
\begin{equation} \label{CURVE4}
 \delta V^{\mu} =   \left( \frac{d{ V^{\mu}}}{d \tau} -  T^{\mu}_{\sigma} \frac{dX^{\sigma}}{d\tau} \right) \delta \tau \quad .
\end{equation}
This equation expresses the change in a vector field due to a translation along a curve parametized with arc length $\tau.$

With {\it{parallel translation}} of $V^{\mu}(\tau)$ along the curve $X^{\nu}(\tau),$ the change (\ref{CURVE4}) vanishes, $\delta V^{\mu}(\tau)$ = 0, and it follows that
\begin{equation} \label{Ptrans1}
     \frac{d{ V^{\mu}}}{d \tau} =  T^{\mu}_{\sigma} \frac{dX^{\sigma}}{d\tau}  \quad .
\end{equation}
The 2nd rank tensor $T^{\mu}_{\sigma}$ may be called a non-linear connection. Conventionally, `parallel translation with connection $\bar{\Gamma}$' implies  an equation much like (\ref{Ptrans1}), but with $T^{\mu}_{\sigma}$ in the form $T^{\mu}_{\sigma}$ = ${\bar{\Gamma}}^{\mu}_{\sigma \rho}V^{\rho} $ and with $\bar{\Gamma}$ independent of $V^{\mu}.$ Such a connection $\bar{\Gamma}$ would be a tensor for the linear spacetime transformations of the Poincar\'{e} group. (When arbitrary functions can act as coordinates, nonlinear transformations can be considered and for these more general transformations the connection need not be and often is not a tensor.)

By parallel translating $V^{\mu}(\tau)$ along a small closed curve $Y^{\nu}(\tau),$ one can find the {\it{round trip change}} in $V^{\mu}.$  For any $\tau$ along $Y,$ one finds, by expanding $T$ and using (\ref{Ptrans1}), to second order in $\delta Y$ that
$$ V^{\mu}(\tau) = \left(V^{\mu}\right)_{0} + \int^{\tau}_{\tau_{0}}\left[ \left(T^{\mu}_{\sigma}\right)_{0} + \left(\frac{\partial{T^{\mu}_{\sigma}}}{\partial{x^{\rho}}}\right)_{0}\left(Y^{\rho}(\tau) - Y^{\rho}(\tau_{0})\right) \right]\frac{dY^{\sigma}}{d\tau} d\tau \quad ,
$$
where $\left(f\right)_{0}$ indicates the function $f$ is evaluated at the initial event $Y(\tau_{0}).$

Let the round trip be completed at $\tau$ = $\tau_{1},$ hence, 
$$ \int^{\tau_{1}}_{\tau_{0}}\frac{dY^{\sigma}}{d\tau} d\tau = Y^{\mu}(\tau_{1}) - Y^{\mu}(\tau_{0}) = 0 \quad .
$$
For a round trip, the above expression reduces to
$$ V^{\mu}(\tau_{1}) = \left(V^{\mu}\right)_{0} + \left(\frac{\partial{T^{\mu}_{\sigma}}}{\partial{x^{\rho}}}\right)_{0} \int^{\tau_{1}}_{\tau_{0}} Y^{\rho} \frac{dY^{\sigma}}{d\tau} d\tau \quad .
$$
The integrals here are directed areas and are antisymmetric in $\rho$ and $\sigma,$ as follows from
$$ \int^{\tau_{1}}_{\tau_{0}}  \frac{d }{d\tau} \left(Y^{\rho} Y^{\sigma}\right) d\tau = \int^{\tau_{1}}_{\tau_{0}} Y^{\rho} \frac{dY^{\sigma}}{d\tau} d\tau + \int^{\tau_{1}}_{\tau_{0}} Y^{\sigma} \frac{dY^{\rho}}{d\tau} d\tau = Y^{\rho}(\tau_{1}) Y^{\sigma}(\tau_{1}) - Y^{\rho}(\tau_{0}) Y^{\sigma}(\tau_{0}) = 0 \quad .
$$
Thus only the antisymmetric part of the partial of $T$ contributes and one is lead to the following expression for the round trip change of a parallel translated vector,
\begin{equation} \label{RoundTrip}
     V^{\mu}(\tau_{1}) - V^{\mu}(\tau_{0}) = \frac{1}{2}\left(\frac{\partial{T^{\mu}_{\sigma}}}{\partial{x^{\rho}}} - \frac{\partial{T^{\mu}_{\rho}}}{\partial{x^{\sigma}}} \right)_{0} \int^{\tau_{1}}_{\tau_{0}} Y^{\rho} \frac{dY^{\sigma}}{d\tau} d\tau \quad .
\end{equation}
Note that the result is consistent with the Theorem stated above.

A {\it{geodesic curve}} is defined to be any curve $X^{\nu}(\tau) $ which has a parallel translated tangent. Thus the vector $V^{\mu}(\tau)$ in (\ref{Ptrans1}) is taken to be the tangent vector,
$$
      V^{\mu} = \frac{dX^{\mu}}{d\tau}  \quad ,
$$
yielding the geodesic equation:
\begin{equation} \label{ARC1}
     \frac{d^2{ X^{\mu}}}{d \tau^{2}} = T^{\mu}_{\sigma} \frac{dX^{\sigma}}{d\tau} \quad .
\end{equation}
Along a geodesic, the 4-acceleration is the scalar product of the connection tensor and the tangent 4-vector.

In the discussion thus far, the way to calculate 4-dimensional distances has not been specified; arc lengths have met just minimal requirements. In what follows two arc length formulas are considered, the simpler one first. 

\section{Flat Space Metric; Electrodynamics}  \label{LorentzForce}

In this Section, the arc length $\bar{\tau}$ is assumed to be measured with the flat space metric, $\eta_{\mu \nu}$ = diag$\{+1,+1,+1,-1\}.$ The square of the arc length $d\bar{\tau}$ of an infinitesimal interval $dx^{\mu}$ is 
\begin{equation} \label{ARC2}
     (d\bar{\tau})^2 =  -\eta_{\alpha \beta} dx^{\alpha}dx^{\beta} = -(dx^{1})^{2}-(dx^{2})^{2}-(dx^{3})^{2}+(dx^{4})^{2}\quad .
\end{equation}
Dividing both sides by the square of the arc length shows that the square of the magnitude of the vector $dx/d{\bar{\tau}}$ is constant,
$$
     -1 =  \eta_{\alpha \beta} \frac{dx^{\alpha}}{d\bar{\tau}}\frac{dx^{\beta}}{d\bar{\tau}} \quad .
$$
If $dx^{\mu}$ is a time interval $dx^{\mu}$ = $\{0,0,0,dt\},$ then $(d\bar{\tau})^2$ = $(dt)^2.$ Thus $d\bar{\tau}$ is the proper time.

Let $X^{\mu}(\bar{\tau})$ be a geodesic. Constant square of the magnitude of the tangent vector $dX^{\mu}/d\bar{\tau}$ and the geodesic equation (\ref{ARC1}) imply that
$$
     0 =  \frac{d}{d\bar{\tau}} \left(\eta_{\alpha \beta} \frac{dX^{\alpha}}{d\bar{\tau}}\frac{dX^{\beta}}{d\bar{\tau}} \right)=  2\eta_{\alpha \beta} \frac{dX^{\alpha}}{d\bar{\tau}}\frac{d^2X^{\beta}}{d\bar{\tau}^2} 
$$ $$
	0 =  \eta_{\alpha \beta} \eta_{\sigma \rho}\frac{dX^{\alpha}}{d\bar{\tau}} \frac{dX^{\sigma}}{d\bar{\tau}} T^{\rho \beta} \quad .
$$
To make the symmetry plain rewrite this as follows
$$0 = \left(\eta_{\sigma \rho} \frac{dX^{\sigma}}{d\bar{\tau}}\right) \left(\eta_{\alpha \beta} \frac{dX^{\alpha}}{d\bar{\tau}} \right)\frac{1}{2}\left(  T^{\rho \beta} + T^{\beta \rho} \right) \quad .
$$
It is sufficient that $T^{\rho \beta}$ be antisymmetric, but one can also argue that for certain physical applications antisymmetry is necessary. 

If the connection tensor is the same for all timelike geodesics, then the tangent vector factors in the above expression are suitably arbitrary so that it is now {\it{necessary}} by the above formula that $T^{\rho \beta}$ be antisymmetric. A `test' particle can move at any sublight speed in any direction at any event in spacetime along the geodesics of a connection tensor. This connection tensor is akin to the force field that is the same for all paths followed by a test particle with various initial velocities.   

Thus consider the set of all timelike vector fields with geodesics in all timelike directions due to the one connection tensor $T^{\rho \beta}.$ The 4-vector parts of the collection of $\psi^{(a)}$s, each labeled by a different value of $a,$ are restricted to be the tangent vectors of geodesics and the tensor part $T^{\alpha \beta}_{0}$ is the same tensor field for any value of $a.$ Since the 4-vector field of any one $\psi^{(a)}$ can have just one value at each spacetime event, it is important that no two geodesics of any one $\psi^{(a)}$ intersect. Hence it may be necessary to consider the field $\psi^{(a)}$ defined over a patch of spacetime rather than over all of spacetime.

Therefore, by symmetry and for the just-described collection of fields $\psi^{(a)},$ it follows that
\begin{equation} \label{LF1}
      T^{\rho \beta} = - T^{\beta \rho} \quad ,
\end{equation}
and the connection tensor $T^{\mu \nu}$ must be antisymmetric.

When the connection tensor $T^{\mu \nu}$ is identified as
\begin{equation} \label{LF3}
     T^{\mu \nu} = - {\frac{e}{m}} F^{\mu \nu}  \quad ,
\end{equation}
the 4-acceleration of a particle of charge $e$ and mass $m$ in an electromagnetic field $F^{\mu \nu}$ is given by the geodesic equation (\ref{ARC1}) since (\ref{ARC1}) is then the electrodynamic force law
\begin{equation} \label{LF4}
     m\frac{d{ V^{\mu}}}{d \bar{\tau}} = e F^{\mu \sigma} V_{\sigma}  \quad ,
\end{equation}
where the 4-velocity vector $dX^{\mu}/d\bar{\tau}$ is denoted here as $V^{\mu}(\bar{\tau}).$ 

The collection of fields $\psi^{(a)}$ is a collection of fields of the time-like tangent vectors to possible particle paths, each $\psi^{(a)}$ in the form
$$
 \psi_{l}^{(a)}(x)  = \pmatrix{ \frac{d{ X^{(a)\mu}}}{d\bar{\tau}} \cr \cr - {\frac{e}{k m}} F^{\alpha \beta} } \quad ,
$$
where `$a$' distinguishes different paths in the one electromagnetic field $F^{\mu \nu}.$ (This form for the $\psi^{(a)}$s assumes the special case $T^{\mu \nu}$ = $k T^{\mu \nu}_{0},$ which occurs for  $C_{2}$ = $-C_{3}$ = $k/2$ and $C_{1}$ = $C_{4}$ = 0 in (\ref{gen4}).)

This section has shown that the general geodesic equation (\ref{ARC1}) specialized to the case of the flat spacetime arc length yields a description of the paths of charged massive particles in flat spacetime.

\section{Gravitational Dynamics}  \label{Gravity}

While the Minkowski metric is a natural choice for particles in flat spacetime, the proper time of actual clocks depends on the distribution of mass in spacetime. Thus, a position dependent metric is indicated by the observed behavior of clocks.\cite{Will} In this section translations along curves with arc length determined by a position dependent metric is considered. Much the same problems arise as occur with the flat spacetime metric in the previous section. 

Let $g_{\alpha \beta}$ be the sufficiently differentiable position dependent metric. The metric is assumed to be symmetric in $\alpha \beta$ and to have an inverse, $g^{\alpha \sigma} g_{\sigma \beta}$ = $\delta^{\alpha}_{\beta},$ where $\delta^{\alpha}_{\beta}$ is the Kronecker delta, i.e. one for $\alpha$ = $\beta$ and zero otherwise. Then the square of the arc length $(d\tau)^2$ of an interval $dx^{\mu}$ between two nearby events is given by 
\begin{equation} \label{METRIC1}
     (d\tau)^2 =  - g_{\alpha \beta} dx^{\alpha}dx^{\beta} \quad .
\end{equation}
One finds that the square of the magnitude of $dx^{\mu}/d\tau$ is constant;
$$
     -1 =  g_{\alpha \beta} \frac{dx^{\alpha}}{d\tau}\frac{dx^{\beta}}{d\tau} \quad .
$$
The arc length $d\tau$ is the {\it{proper time}}. When $g_{\alpha \beta}$ = $\eta_{\alpha \beta},$ this proper time reduces to $d\bar{\tau}$.

 Let $X^{\mu}(\tau)$ be a geodesic with the $\tau$ in (\ref{METRIC1}) as arc length. Then the square of the magnitude of the tangent vector $dX^{\mu}/d\tau$ is constant, and one finds that 
$$
     0 =  \left( \frac{dg_{\alpha \beta}}{d\tau}\frac{dX^{\alpha}}{d\tau}\frac{dX^{\beta}}{d\tau} +  g_{\alpha \beta} \frac{d^2 X^{\alpha}}{d\tau^2}\frac{dX^{\beta}}{d\tau} +  g_{\alpha \beta} \frac{d X^{\alpha}}{d\tau}\frac{d^2 X^{\beta}}{d\tau^2}\right) \quad .
$$ 
The geodesic equation, (\ref{ARC1}), implies that
$$
	0 = \left( \frac{dg_{\alpha \beta}}{d\tau} + g_{\sigma \beta} T^{\sigma}_{\alpha} + g_{\alpha \sigma} T^{\sigma}_{\beta}\right) \frac{dX^{\alpha}}{d\tau}\frac{dX^{\beta}}{d\tau}  \quad .
$$

In order to have an additional tangent vector $ (dX^{\gamma}/d\tau)$ factor in the above expression, express the derivative of the metric with respect to arc length as  
$$
     \frac{d g_{\alpha \beta}}{d\tau} =  \frac{\partial{g_{\alpha \beta}}}{dx^{\gamma}} \frac{dX^{\gamma}}{d\tau} \quad ,
$$
Also, define fields $U^{\sigma}_{\alpha}(x)$ and $\Gamma^{\sigma}_{\alpha \gamma}(x)$ so that
 \begin{equation} \label{GEOD4}
       T^{\sigma}_{\alpha} = g_{\alpha \rho}U^{\sigma \rho} + \Gamma^{\sigma}_{\alpha \gamma} \frac{dX^{\gamma}}{d\tau}  \quad ,
\end{equation}
and require that $U^{\sigma \rho}(x)$ satisfies the following equation
 \begin{equation} \label{GEOD5a}
       \left(g_{\sigma \beta} g_{\alpha \rho}U^{\sigma \rho} + g_{\alpha \sigma} g_{\beta \rho}U^{\sigma \rho}\right) \frac{dX^{\alpha}}{d\tau}\frac{dX^{\beta}}{d\tau} = 0 \quad .
\end{equation}
As long as the tangent vector $ (dX^{\gamma}/d\tau)$ never vanishes, there exist 64-component solutions $\Gamma^{\sigma}_{\alpha \gamma}$ to the 16 equations (\ref{GEOD4}), so there is no loss of generality in rewriting the connection.  

By (\ref{GEOD4}) and (\ref{GEOD5a}), $ (dX^{\gamma}/d\tau)$ factors out of the expression above, giving
 \begin{equation} \label{GEOD5b}
       0 = \left( \frac{\partial{g_{\alpha \beta}}}{\partial{x^{\gamma}}} + g_{\sigma \beta} \Gamma^{\sigma}_{(\alpha \gamma)} + g_{\alpha \sigma} \Gamma^{\sigma}_{(\beta \gamma)} \right) \frac{dX^{\alpha}}{d\tau}\frac{dX^{\beta}}{d\tau} \frac{dX^{\gamma}}{d\tau} \quad ,
\end{equation}
where $\Gamma^{\alpha}_{(\beta \gamma)} \equiv$  $(\Gamma^{\alpha}_{\beta \gamma} + \Gamma^{\alpha}_{\gamma \beta})/2$ is the `symmetric part' of $\Gamma.$ Only the symmetric part contributes because of the symmetry in the tangent factors. The antisymmetric part $\Gamma^{\alpha}_{[\beta \gamma]} \equiv$  $(\Gamma^{\alpha}_{\beta \gamma} - \Gamma^{\alpha}_{\gamma \beta})/2$ is often called the {\it{torsion}}.

It is sufficient, but not necessary that the expressions in parentheses in (\ref{GEOD5a}) and (\ref{GEOD5b}) vanish. As in Section \ref{LorentzForce}, one can argue that for conventional applications to particle motion the expression in parentheses must necessarily vanish. Assume that `test' particles find themselves in the same conditions set by fields $U^{\sigma \rho}$ and $\Gamma^{\sigma}_{(\alpha \gamma)}.$ The same fields are thereby present for a test particle that may have any subluminal velocity at any spacetime event. The range of allowed velocities is assumed to be 4-dimensional for the lightcone determined by the position dependent metric. Hence, for such a case based on physical expectations, the tangent vectors are suitably arbitrary and one concludes that the expressions in parentheses in (\ref{GEOD5a}) and (\ref{GEOD5b}) vanish necessarily.

To accomplish this, assume that the tensor field $\Gamma^{\beta}_{\sigma \gamma}$ is the same for all the fields of geodesic tangent vectors $d{ X^{(a)\mu}}/d\tau$ in a collection of fields $\psi_{l}^{(a)}(x).$ Thus each $\psi_{l}^{(a)}(x)$ in the collection has the form
$$
 \psi_{l}^{(a)}(x)  = \pmatrix{ v^{(a)\mu} \cr \cr k^{-1} \eta^{\alpha \sigma}(g_{\sigma \gamma} U^{\beta \gamma} + \Gamma^{\beta}_{\sigma \gamma} v^{(a)\gamma}) } \quad .
$$
where the tensor fields $U^{\beta \gamma}$ and  $\Gamma^{\beta}_{\sigma \gamma}(x)$ are the same for all $\psi_{l}^{(a)}(x)$ in the collection. [The form for the $\psi$s displayed here assumes the special case $T^{\mu \nu}$ = $k T^{\mu \nu}_{0},$ which occurs for  $C_{2}$ = $-C_{3}$ = $k/2$ and $C_{1}$ = $C_{4}$ = 0 in (\ref{gen4}).]

A collection of global fields $\psi_{l}^{(a)}(x)$ may not exist, since geodesics can intersect. So, to keep  both the 4-vector and tensor parts of $\psi_{l}^{(a)}(x)$  single valued, one may define a collection $\psi_{l}^{(a)}(x)$ in one patch and cover spacetime with overlapping patches. 

In particular, by (\ref{GEOD5b}) with suitably arbitrary tangent vectors, one has
 \begin{equation} \label{affine}
       0 =  \frac{\partial{g_{\alpha \beta}}}{\partial{x^{\gamma}}} + g_{\sigma \beta} \Gamma^{\sigma}_{(\alpha \gamma)} + g_{\alpha \sigma} \Gamma^{\sigma}_{(\beta \gamma)}  \quad .
\end{equation}

Combining (\ref{affine}) with the two equations found by permuting the indices $\alpha \beta \gamma$ in (\ref{affine}), and since the metric is symmetric and has an inverse, one finds that
 \begin{equation} \label{Christoffel}
       \Gamma^{\alpha}_{(\beta \gamma)} = -\frac{g^{\alpha \sigma} }{2} \left( \frac{\partial{g_{\beta \sigma}}}{\partial{x^{\gamma}}} + \frac{\partial{g_{\gamma \sigma}}}{\partial{x^{\beta}}} - \frac{\partial{g_{\beta \gamma}}}{\partial{x^{\sigma}}} \right)  \quad ,
\end{equation}
which is the negative of the Christoffel connection familiar from general relativity. Thus $\Gamma^{\alpha}_{(\beta \gamma)}$ is determined by the metric $g_{\alpha \beta}.$
 
With the connection tensor in (\ref{GEOD4}), the geodesic equation (\ref{ARC1}) gives
\begin{equation} \label{GEOD8}
       \frac{d^2{ X^{(a)\mu}}}{d\tau^2} = U^{(a)\sigma}_{\alpha}\frac{d{ X^{(a)\alpha}}}{d\tau} + \Gamma^{\mu}_{(\alpha \gamma)} \frac{d{ X^{(a)\alpha}}}{d\tau} \frac{dX^{(a)\gamma}}{d\tau}   \quad ,
\end{equation}
where the tensor field $U^{(a)\sigma}_{\alpha}(x)$ satisfies (\ref{GEOD5a}).

For the collection of suitably arbitrary tangent vectors $d{ X^{(a)\alpha}}/d\tau$ and by the symmetry apparent in (\ref{GEOD5a}), it follows that $U^{\alpha \beta}$ is necessarily antisymmetric. Suppose $U^{\alpha \beta}$ is chosen to be proportional to the electromagnetic field $F^{\rho \omega}(x),$ 
$$
       U^{\alpha \beta} = \frac{e}{m} F^{\alpha \beta}   \quad ,
$$
where $e$ is the charge and $m$ is the mass of the particles of the fields $\psi^{(a)}_{l}(x).$ 
Then the connection tensor is given by
 \begin{equation} \label{GEOD10}
       T^{(a)\nu}_{\alpha} = \frac{e}{m}g_{\alpha \sigma}F^{\nu \sigma} + \Gamma^{\nu}_{\alpha \sigma} \frac{dX^{(a)\sigma}}{d\tau}  \quad ,
\end{equation}
where $F^{\nu \sigma}$ is the electromagnetic field and the symmetric part of $-\Gamma^{\nu}_{\alpha \sigma},$ $-\Gamma^{\nu}_{(\alpha \sigma)},$ is the Christoffel connection of the metric $g_{\alpha \beta}.$  
Each field in the collection of fields $\psi^{(a)}$ now has the form
$$
 \psi_{l}^{(a)}(x)  = \pmatrix{ v^{(a)\mu} \cr \cr k^{-1} \eta^{\alpha \delta}(-\frac{e}{m}g_{\delta \kappa}F^{\kappa \beta} + \Gamma^{\beta}_{\delta \gamma} v^{(a)\mu} )} \quad ,
$$
when $T^{\alpha \beta}$ = $k T^{\alpha \beta}_{0}$ with momentum matrix constants $C_{2}$ = $-C_{3}$ = k/2 and $C_{1}$ = $C_{4}$ = 0, where the constant $k$ has units of inverse distance. 

With the connection tensor in (\ref{GEOD10}), the geodesic equation (\ref{GEOD8}) becomes
\begin{equation} \label{GEOD11}
       \frac{d^2{ X^{(a)\mu}}}{d\tau^2} = \frac{e}{m}g_{\alpha \rho}F^{\mu \rho }\frac{d{ X^{(a)\alpha}}}{d\tau} + \Gamma^{\mu}_{(\alpha \gamma)} \frac{d{ X^{(a)\alpha}}}{d\tau} \frac{dX^{(a)\gamma}}{d\tau}   \quad ,
\end{equation}
which is the force law obtained in general relativity for the motion of particles of charge $e$ and mass $m$ in the electromagnetic field $F^{\mu \nu}$ with infinitesimal proper times and distances determined by the metric $g_{\alpha \beta}$.

The round trip deviation  for the connection tensor (\ref{GEOD10}) can be found following the steps that lead to (\ref{RoundTrip}). Let vector $V^{\mu}$ be parallel translated along a closed curve $Y^{\mu}(\tau)$ with coincident endpoints $Y^{\mu}(\tau_{1})$ = $Y^{\mu}(\tau_{0}).$ Along $Y$ the connection tensor $T^{\mu}_{\sigma}$ is a function of $\tau$ given by the expression

$$T^{\mu}_{\sigma}  = \frac{e}{m} F^{\mu}_{\sigma} +  \Gamma^{\mu}_{\sigma \kappa} V^{\kappa} \quad ,
$$ 
where the label $a$ is dropped and $ F^{\mu}_{\sigma} \equiv$ $g_{\sigma \rho}F^{\mu \rho }$ . Note that the torsion $\Gamma^{\mu}_{[\sigma \kappa]}$ contributes since $\Gamma^{\mu}_{\sigma \kappa}$ has not been assumed to be symmetric.

For any $\tau$ along $Y,$ one finds by the parallel translation rule (\ref{Ptrans1}),  to second order in $\delta Y,$ that
$$ V^{\mu}(\tau) - V^{\mu}(\tau_{0})  = \hspace{12cm}
$$ 
$$  \int^{\tau}_{\tau_{0}}\left\{ \left(T^{\mu}_{\sigma}\right)_{0} +  \left[\frac{e}{m} \left(\frac{\partial{F^{\mu}_{\sigma}}}{\partial{x^{\rho}}} +\Gamma^{\mu}_{\sigma \kappa} F^{\kappa}_{\rho}\right) + \left(\frac{\partial{\Gamma^{\mu}_{\sigma \kappa}}  }{ \partial{x^{\rho} }} + \Gamma^{\mu}_{\sigma \eta} \Gamma^{\eta}_{\rho \kappa} \right) V^{\kappa}\right]_{0}      
\left(Y^{\rho}(\tau) - Y^{\rho}(\tau_{0})\right) \right\}\frac{dY^{\sigma}}{d\tau} d\tau 
$$
where $\left(f\right)_{0}$ indicates the function $f$ is evaluated at the initial event $Y(\tau_{0}).$

Let the curve be closed at $\tau_{1}$ so that $\int^{1}_{0} dY^{\sigma}$ = 0 and that the directed area $\int^{1}_{0} Y^{\rho} dY^{\sigma}$ is antisymmetric in $\rho \sigma,$ one finds that
\begin{equation} \label{Curvature}
     V^{\mu}(\tau_{1}) - V^{\mu}(\tau_{0}) = \frac{1}{2}\left[\frac{e}{m}\left( \frac{DF^{\mu}_{\sigma}}{\partial x^{\rho}} - \frac{DF^{\mu}_{\rho}}{\partial x^{\sigma}}\right) +  R^{\mu}_{\sigma \kappa \rho} V^{\kappa} \right] \int^{\tau_{1}}_{\tau_{0}} Y^{\rho} \frac{dY^{\sigma}}{d\tau} dt \quad ,
\end{equation}
where the `covariant derivatives' of $F$ are defined as follows
$$ \frac{DF^{\mu}_{\sigma}}{\partial x^{\rho}} \equiv \frac{\partial{F^{\mu}_{\sigma}}}{\partial x^{\rho}} + F^{\omega}_{\sigma} \Gamma^{\mu}_{\rho \omega} - F^{\mu}_{\alpha} \Gamma^{\alpha}_{\rho \sigma} \quad ,
$$
and the curvature tensor $R^{\mu}_{\sigma \kappa \rho}$ is given by
$$R^{\mu}_{\sigma \kappa \rho} \equiv \frac{\partial{\Gamma^{\mu}_{\sigma \kappa}}}{\partial{x^{\rho}}} - \frac{\partial{\Gamma^{\mu}_{\rho \kappa}}}{\partial{x^{\sigma}}} +\Gamma^{\mu}_{\sigma \eta}\Gamma^{\eta}_{\rho \kappa} - \Gamma^{\mu}_{\rho \eta}\Gamma^{\eta}_{\sigma \kappa} \quad .
$$

Note that the round trip deviation is inhomogeneous in the translated vector $V^{\kappa},$ so there is a deviation even for an initially null vector.
When $g_{\alpha \beta}$ is the constant flat space metric $\eta_{\alpha \beta},$ the round trip deviation reduces to $e/m$ times the gradient of the electromagnetic flux through the loop $Y^{\mu}(\tau),$ see the problem set in Appendix \ref{A}.

By (\ref{Christoffel}), when the torsion vanishes, the curvature tensor $R^{\mu}_{\sigma \kappa \rho}$ is the Riemann-Christoffel curvature tensor of the position dependent metric $g_{\alpha \beta}.$ And, with a null electromagnetic field, the round trip deviation (\ref{Curvature}) is just the same as in general relativity.\cite{WeinbergPtrans}

\section{Mapping Flat Spacetime}  \label{Survey}

If, far from gravitational sources, the proper time of a particle is sufficiently accurately determined by the Minkowski metric, then the conceptual problem of laying out markers for the spacetime coordinates is solved by using electrically neutral, nonmagnetic rulers and clocks. These measuring devices would behave as needed for constructing the flat space coordinate system $x^{\mu}.$ This construction of flat spacetime reference frames is described in detail in many special relativity texts.

But, closer to gravitational sources, proper time must be calculated with the position dependent gravitational metric. The observed red shift and other observed behavior of matter, complicates the conceptual construction of the flat spacetime coordinates $x^{\mu}.$  Remember, the dynamic laws obtained here as geodesic equations are derived from the properties of representations of the symmetries of flat spacetime, not curved spacetime. 

So, what measurements, in principle, could determine the coordinates $x^{\mu}$ of an event in flat spacetime? One could try to work backwards from the geodesic equation by deducing $X(\tau)$ from observations of paths of objects in the gravitational field. But in principle the process is flawed. The objection is that general relativistic metrics determined by field equations can be altered by curvilinear coordinate transformations and still obey the field equations. Given any one metric, another could be generated by a curvilinear coordinate transformation that approaches the identity far away from the source. Many equally qualified candidates would produce different sets of coordinates $x^{\mu}.$ One could perhaps select by brute force the harmonic solution \cite{WeinbergHARM} to the field equations, but why would that be justified? No, it seems that observing the behavior of ordinary matter in gravitational fields would not uniquely yield the underlying flat spacetime.

Perhaps whimsically and fortuitously, there could be a form of matter that has charge and has mass, but that follows the flat spacetime geodesics whose arc length is measured with the Minkowski metric. Such matter would not fall and the equivalence principle would be violated, but the existence of such material would be a great convenience to surveying the flat spacetime $x^{\mu},$ especially if clocks and rulers could be fashioned from the material. Naturally, it may be that these as-yet-unseen speculative forms of matter seemingly needed for mapping the flat spacetime $x^{\mu}$ may in fact not exist. So it may be impossible to describe a sequence of measurements that can be made using real particles that can map the flat spacetime whose symmetries are the basis for the connections' existence that underlies the derivation of realistic geodesic particle motion presented in this article.

\appendix

\section{Problems} \label{A}

\vspace{0.3cm}
\noindent 1. Consider the momentum matrices in (\ref{gen4}). (a) For what values of $C_{i}$ is $T^{\mu \nu}$ symmetric in $\mu \nu$? (b) Antisymmetric? (c) The trace of a tensor, $\bar{T} \equiv$ $\bar{T}^{11} + \bar{T}^{22} + \bar{T}^{33} - \bar{T}^{44}$ is an invariant. Find constants $C_{i}$ such that $T^{\mu \nu}$ is $\eta^{\mu \nu}$ times the trace $T_{0}$ of the tensor part of $\psi.$ How does this relate to the spin $(1/2,1/2) \oplus(0,0)$ coordinate representation of the Poincar\'{e} group?

\vspace{0.3cm}
\noindent 2. Rewrite the geodesic equation with a new arc length $s$. That is, simplify the equation resulting from substituting the function $\tau(s)$ for $\tau$ in (\ref{ARC1}). In particular, try $\tau(s)$ =  $c_0 + c_1 s,$ where $c_{i}$ are constants.

\vspace{0.3cm}
\noindent 3. Using the expressions in Section \ref{psi}, construct the 20x20 matrix generator $J^{12}$ of rotations in the 12-plane (the $xy$-plane). Also construct the matrix generator of boosts in the 3 direction, $J^{34}.$  To do this convert the 16 values of the double indices ${\gamma \delta}$ and ${\epsilon \xi}$ to single indices that range from 5 to 20 because 1 to 4 is reserved for the vector part. (One choice of ordering is $\{12,13,14,23,24,34,$ $11,22,33,44,21,31,41,32,42,43 \},$ so that a single index of $i$ = $4+5$ = 9 indicates the double index ${\gamma \delta}$ = 24.)

\vspace{0.3cm}
\noindent 4. (a) Show that the 11-block matrices $J^{\alpha \beta}_{11}$ in (\ref{gen1}) satisfy the Poincar\'{e} algebra commutation rule
$$  [J^{\mu \nu}_{11},J^{\rho \sigma}_{11}] = i [ \eta^{\mu \rho} J^{\nu \sigma}_{11} -  \eta^{\mu \sigma} J^{\nu \rho}_{11} -  \eta^{\nu \rho} J^{\mu \sigma}_{11} +  \eta^{\nu \sigma} J^{\mu \rho}_{11}] \quad ,
$$
where $[J^{\mu \nu}_{11},J^{\rho \sigma}_{11}]^{\alpha}_{\gamma}$ = $(J^{\mu \nu}_{11})^{\alpha}_{\beta}  (J^{\rho \sigma}_{11})^{\beta}_{\gamma} - (J^{\rho \sigma}_{11})^{\alpha}_{\beta}  (J^{\mu \nu}_{11})^{\beta}_{\gamma}$  .

(b) Show that the momentum-angular momentum commutation rule is satisfied by the expressions in Section \ref{psi}. The [P,J] rule is trivial except for the 12-block,
$$[P^{\mu},J^{\rho \sigma}]_{12} = (P^{\mu}_{12})^{\nu}_{\alpha \beta}(J^{\rho \sigma}_{22})^{\alpha \beta}_{\gamma \delta} - (J^{\rho \sigma}_{11})^{\nu}_{\sigma} (P^{\mu}_{12})^{\sigma}_{\gamma \delta} = -i \left( \eta^{\mu \rho} P^{\sigma}_{12} -  \eta^{\mu \sigma} P^{\rho}_{12}\right) \quad .
$$

\vspace{0.3cm}
\noindent 5.  Suppose the closed curve $Y^{\mu}(\tau)$ is a small loop in the 12-plane and that the metric is the constant metric $\eta_{\alpha \beta}.$ Show that the time component of the round trip deviation is proportional to the electromotive force developed around the loop when the electromagnetic field is a time dependent magnetic field in the 3-direction. (Hint: Use a Maxwell equation.)

\vspace{0.3cm}
\noindent 6.  Show that, if the round trip deviation (\ref{Curvature}) vanishes for all small closed curves containing an event $x^{\mu}$, then one has $ R_{\rho \sigma } V^{\sigma}$ = $(e/m) j_{\rho}$ at the event $x^{\mu},$ where the Ricci tensor is $ R_{\rho \sigma} \equiv$  $ R^{\alpha}_{\rho \alpha \sigma}$ and $j_{\rho}$ is the electromagnetic current 4-vector, $D{F^{\sigma}_{\rho}}/{\partial{x^{\sigma}}}$ = $j_{\rho},$ by one of the Maxwell equations.

\end{document}